\documentclass[letter]{aa} 

\usepackage{amsmath}
\usepackage{graphicx}
\usepackage{adjustbox}
\usepackage{longtable}
\usepackage{geometry} 
\usepackage{array}
\usepackage{caption} 
\usepackage{subcaption} 
\usepackage{lscape} 
\usepackage{hyperref}

\usepackage{txfonts}
\begin{document}

   \title{An "alien" called Oosterhoff dichotomy?}

   \author{E. Luongo
          \inst{1}
          \and
          V. Ripepi\inst{1}
          \and
          M. Marconi\inst{1}
          \and
          Z. Prudil\inst{2}
          \and
          M. Rejkuba\inst{2}
          \and
          G. Clementini\inst{3}
          \and
          G. Longo\inst{4}
          }

   \institute{INAF - Osservatorio Astronomico di Capodimonte, Via Moiariello 16, I-80131 Naples, Italy.\\
              \email{emanuela.luongo@inaf.it}
         \and
             European Southern Observatory, Karl-Schwarzschild-Strasse 2, 85748 Garching bei München, Germany
         \and
            INAF, Osservatorio di Astrofisica e Scienza dello Spazio di Bologna, via Piero Gobetti 93/3, 40129 Bologna, Italy
         \and
            Department of Physics, University Federico II, via Cinthia 6, 80126 Napoli, Italy\\
            }

   \date{Received September 15, 1996; accepted March 16, 1997}

 
  \abstract
   {}
   {In this letter we investigate the origin of the Oosterhoff dichotomy, considering recent discoveries related to several ancient merging events of external galaxies with the Milky Way (MW). In particular, we aim to clarify if the subdivision in Oosterhoff type of Galactic Globular Clusters (GGCs) and field RR Lyrae (RRLs)  could be traced back to one or more ancient galaxies that merged with the MW in its past.}
   {To this purpose, we first explored the association of GGCs with the past merging events according to different literature studies. Subsequently we compiled positions, proper motions and radial velocity for 10,138 field RRLs variables from the  $Gaia$  Data Release 3. To infer the distances, we adopted the $M_G$--[Fe/H] relation, with [Fe/H] values  estimated through empirical relationships involving the individual periods and Fourier parameters. We then calculated the orbits and the integrals of motions (IoM) using the Python library Galpy for the whole sample. By comparing the location of the field RRLs in the energy-angular momentum diagram with that of the GGCs we assign their likely origin. Finally, we discriminate from the $Gaia$ G-band light curves the Oosterhoff type of our sample of RRL stars based on their location in the Bailey diagram.}
   {The analysis of the Bailey diagrams for Galactic RRLs stars and GGCs associated with \textit{In-Situ} vs \textit{Accreted} halo origin shows remarkable differences. The \textit{In-Situ} sample displays a wide range of metallicities with a continuous distribution and no sign of Oosterhoff dichotomy. Conversely, the \textit{Accreted} RRLs clearly shows the Oosterhoff dichotomy and a significantly smaller dispersion in metallicity.} 
    {Our results suggest that the Oosterhoff dichotomy was imported into the MW by the merging events that shaped the Galaxy.}


   \maketitle
%

\section{Introduction}
\label{Sec:Introduction}
RRLs are important standard candles and excellent tracers of the old populations (age $>$ 10 Gyr) in the MW and the Local Group \citep[][]{Smith1995,catelan2015pulsating}. One of the open problems concerning RRLs is the Oosterhoff dichotomy, namely the sub-division of GGCs into two distinct groups according to the average period $<P_{ab}>$ of their fundamental mode RRL variables \citep[RRab,][]{oosterhoff1939some}. In particular, in some GGCs containing RRLs, $<P_{ab}>$ is about 0.55 days while in others, it is close to 0.65 days, with a period gap between these two groups, named Oosterhoff I (OoI) and Oosterhoff II (OoII), respectively.\\
Among the various explanations suggested in the past for this dichotomy \citep[e.g.,][for a recent review]{Fabrizio2021}, \citet{sandage1981evidence} proposed that the dichotomy originated from an intrinsically higher luminosity of OoII RRab at a fixed temperature (assuming similar mass), resulting in longer periods. However, this  implies an anticorrelation between metallicity and helium, since OoII GGCs are generally more metal-poor than the OoI ones and only a significant He enrichment can justify the assumed OoII RRLs overluminosity \citep[e.g.,][]{Bencivenni1991}. 
Alternatively, \citet{van1973two} hypothesized that in the "OR" region of the instability strip \footnote{The OR zone is located between the blue edge of the fundamental mode and the red edge of the first overtone mode. In this region, the RRLs can pulsate in both modes, sometimes simultaneously as double pulsators.} the RRLs found within OoI and OoII GGCs pulsate in the fundamental and first overtone modes, respectively. This was later supported by theoretical work by \citet{bono1997nonlinear}. In this case, OoI RRab would extend to higher temperature and thus in turn have a shorter $<P_{ab}>$\footnote{The period decreases as the effective temperature increases.} than those in OoII.
Recently, \citet{Fabrizio2019} found that their spectroscopic sample of field RRLs plotted in the Bailey diagram show a continuous variation when moving from the metal-poor ([Fe/H]  $\sim -$3.0 dex) to the metal-rich ([Fe/H] $\sim 0$ dex) regime. They concluded that the smooth transition in the peak of the period distribution as a function of the metallicity proves that the Oostehoff dichotomy is not present in the field and its occurrence in GGCs is due to the lack of metal-intermediate clusters hosting RRLs. 
However, the recent $Gaia$ Data Release 3 \citep[DR3][]{prusti2016gaia,GaiaVallenari2023,Clementini2023} showed a clear separation between field OoI and OoII RRL locations in the Bailey diagram, thus 
suggesting that the Oosterhoff dichotomy is still an open problem.


Moreover, \citet{catelan2009horizontal} noted that the gap between the two types of Oosterhoff, when plotting the average RRab period as a function of their metallicity, is filled by GCs in MW's satellite galaxies, identifying them as intermediate Oosterhoff types. This suggests that the Oosterhoff dichotomy might be useful for understanding the mechanisms underlying the formation of the Galactic halo.\\
Recently, the \textit{Gaia} data releases DR2 and DR3 \citet{GaiaBabu2018,Helmi2018a,helmi2020streams}, led to discovery of several ancient merging events in the MW's past, most notably Gaia-Enceladus, Sequoia, Kraken and others, with new insights into the formation and evolution of our Galaxy \citep[e.g.,][and references therein]{GaiaBabu2018,Helmi2018a,helmi2020streams}.
These discoveries were possible because the remnants of the progenitor galaxies occupy the same region in the space of the integrals of motion (IoM, see following sections) as predicted by the numerical simulations of  \citet{helmi2000mapping}. In particular, \citet{Massari2019}, by studying the IoM of the GGCs, were able to associate a significant number of them to different progenitor galaxies. 
This discovery led us to hypothesize a possible connection between the ancient merging episodes that shaped the Galactic halo and the Oosterhoff dichotomy. Is it possible that the Oosterhoff dichotomy got "imported" into the MW?
In the following, we try to answer this question using a sample of GGCs and field RRLs for which it was possible to calculate the IoM and determine their origin.


\section{Observed sample of GGCs}
\label{Sec:Observed_sample_of_GGCs}
We adopted the GGCs inventory and relative fundamental parameters listed by Baumgardt and collaborators\footnote{\href{https://people.smp.uq.edu.au/HolgerBaumgardt/globular/}{https://people.smp.uq.edu.au/HolgerBaumgardt/globular/}} 
\citep[B\&Co hereafter, see also][]{Baumgardt2021,Vasiliev2021}. This database reports accurate astrometric, kinematic and structural data for GGCs, largely based on the results of the $Gaia$ satellite. 
Concerning the possible association of each GGC to a past merging event, we relied on the classification originally proposed by \citet{Massari2019} and further refined by \citet{callingham2022chemo}. In the latter work, almost all GGCs are associated with a progenitor by using the IoM and the actions and adopting a complex Machine Learning technique. The 8 identified progenitors are listed in Table~\ref{tab:prog}.
To further explore the association with ancient merging episodes, we also considered the separation into \textit{In-Situ} and \textit{Accreted} clusters determined by \citet{belokurov2024situ} for 158 of the GGCs listed by B\&Co, using the [Al/Fe] ratio \footnote{Aluminium is produced by type II supernovae and only partially by AGB stars, therefore its abundance is sensitive to different chemical enrichment histories \citep[][]{hawkins2015using}}.

\citet{belokurov2024situ} reported the result of this Aluminium-based classification on the ($E$, $L_z$) plane showing that they can identify the GGCs populations with distinct spatial, kinematic and chemical abundance distributions. \\
The last piece of information we gather from the literature concerns the Oosterhoff type, which is found for only 47 GGCs from  \citet{stobie1971difference} and \citet{van2011some}. \\
Finally, we computed the IoM for each GGC. Assuming an axisymmetric potential for the MW, the commonly considered IoM are the energy $E$, the angular momentum component along the z-axis perpendicular to the Galactic plane $L_z$, and the modulus of the angular momentum vector lying in the Galactic plane, $L_{\bot} = \sqrt{{L_x}^2 + {L_y}^2}$. 
However, the action quantities ($J_r$, $J_{\phi}$, $J_z$), being adiabatic invariants, are better conserved for the sample of Gaia DR3, which extends considerably farther than the previous release \citep[][]{helmi2020streams}. 
In the following, we use the quantities ($E$, $L_z$) to represent the considered GGCs in the same plane as used in the classification method papers \citep{belokurov2024situ, callingham2022chemo}.
In addition, the action quantities ($J_r$, $J_{\phi}$, $J_z$) will be used to classify the field RRLs sample with the {\tt RandomForestClassifier} as shown in Sect.~\ref{Sec:Random_Forest_to_classify_field_RRLs}). 

To calculate the IoM and the actions we adopted the Python library {\tt Galpy} \citep[][]{bovy2015galpy} by integrating the orbits of the targets for a period of 3 Gyrs. The input 6-dimensional parameters (right ascension $RA$, declination $DEC$, distance $d$, proper motion right ascension $\mu_{\rm RA}$ and declination $\mu_{\rm DEC}$, radial velocity RV) for the GGCs were taken from the B\&Co database. These quantities allow us to compute the IoM with {\tt Galpy} providing a potential for the MW, which is, in our case, the axisymmetric \textit{McMillan2017} potential \citep[for details see][]{mcmillan2016mass}. 
The sample of GGCs used in this work together with all the information needed to calculate the IoM and the actions, as well as the classification in terms of \textit{In-Situ} or \textit{Accreted}, are reported in Table~\ref{Tab:5GGCs_dataset}.

\begin{figure*}[htbp]
    \centering
    \vbox{
    \hbox{
    \includegraphics[width=0.45\textwidth, height=0.45\textheight, keepaspectratio]{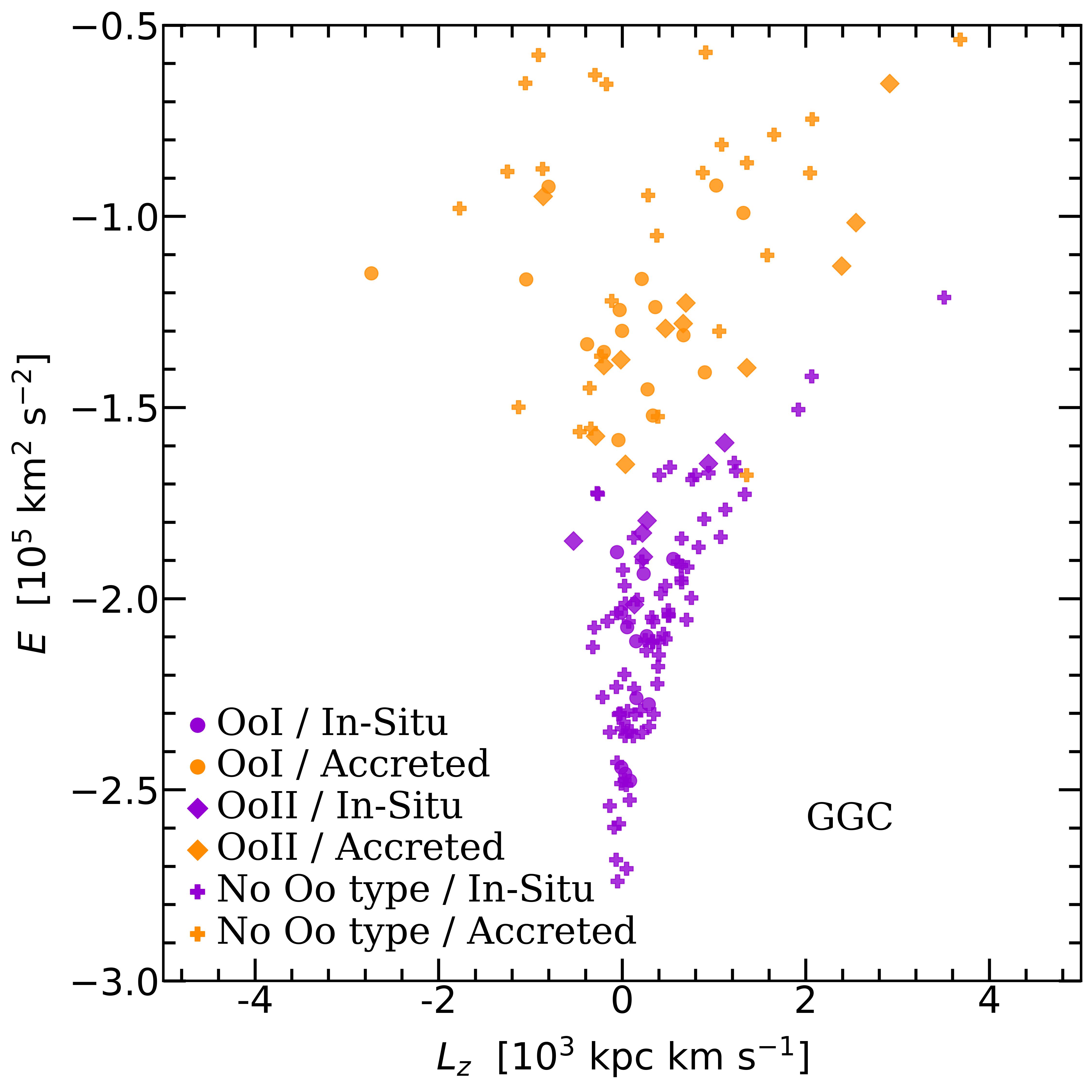}
    \includegraphics[width=0.45\textwidth, height=0.45\textheight, keepaspectratio]{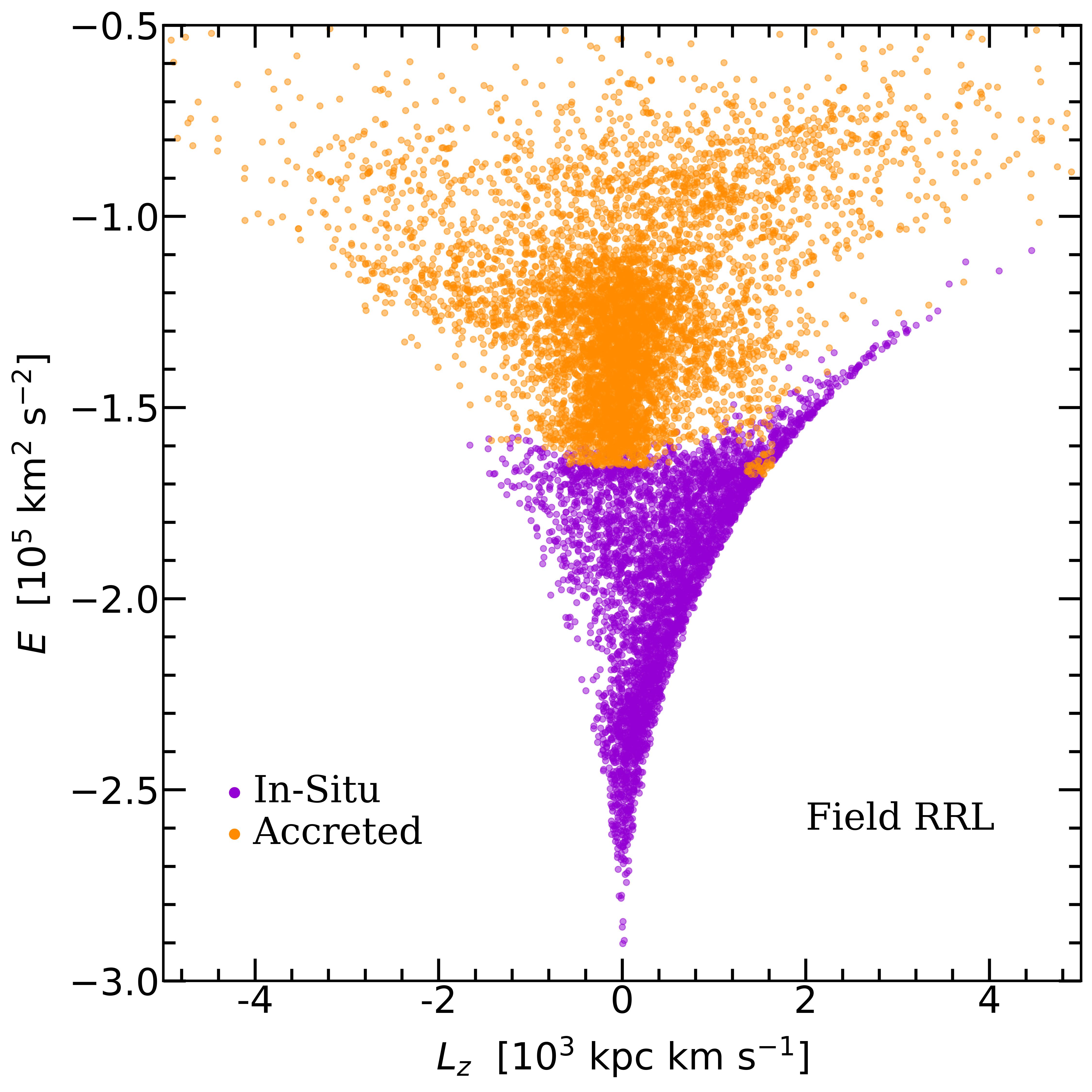}}
    \hbox{
    \includegraphics[width=0.45\textwidth, height=0.45\textheight, keepaspectratio]{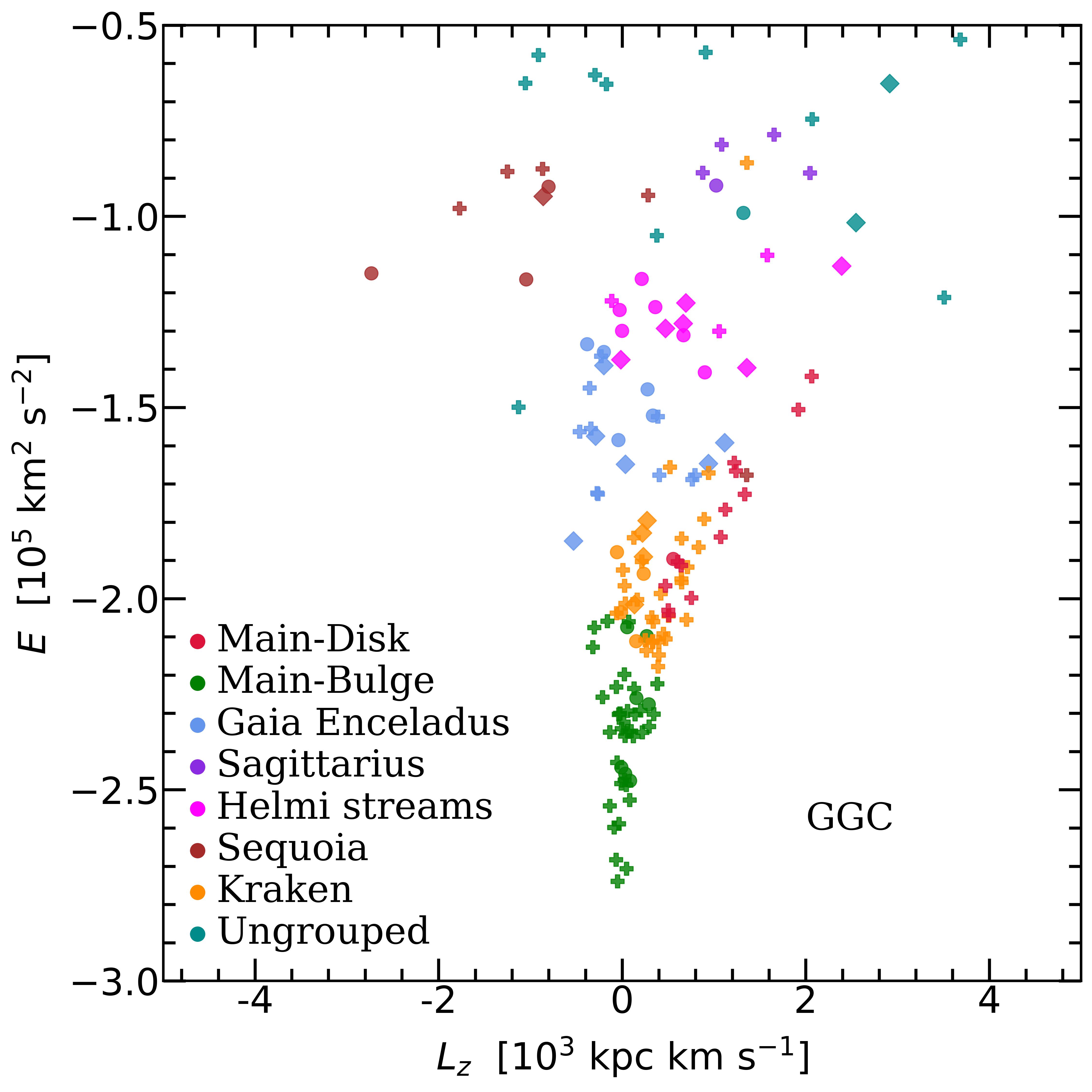}
    \includegraphics[width=0.45\textwidth, height=0.45\textheight, keepaspectratio]{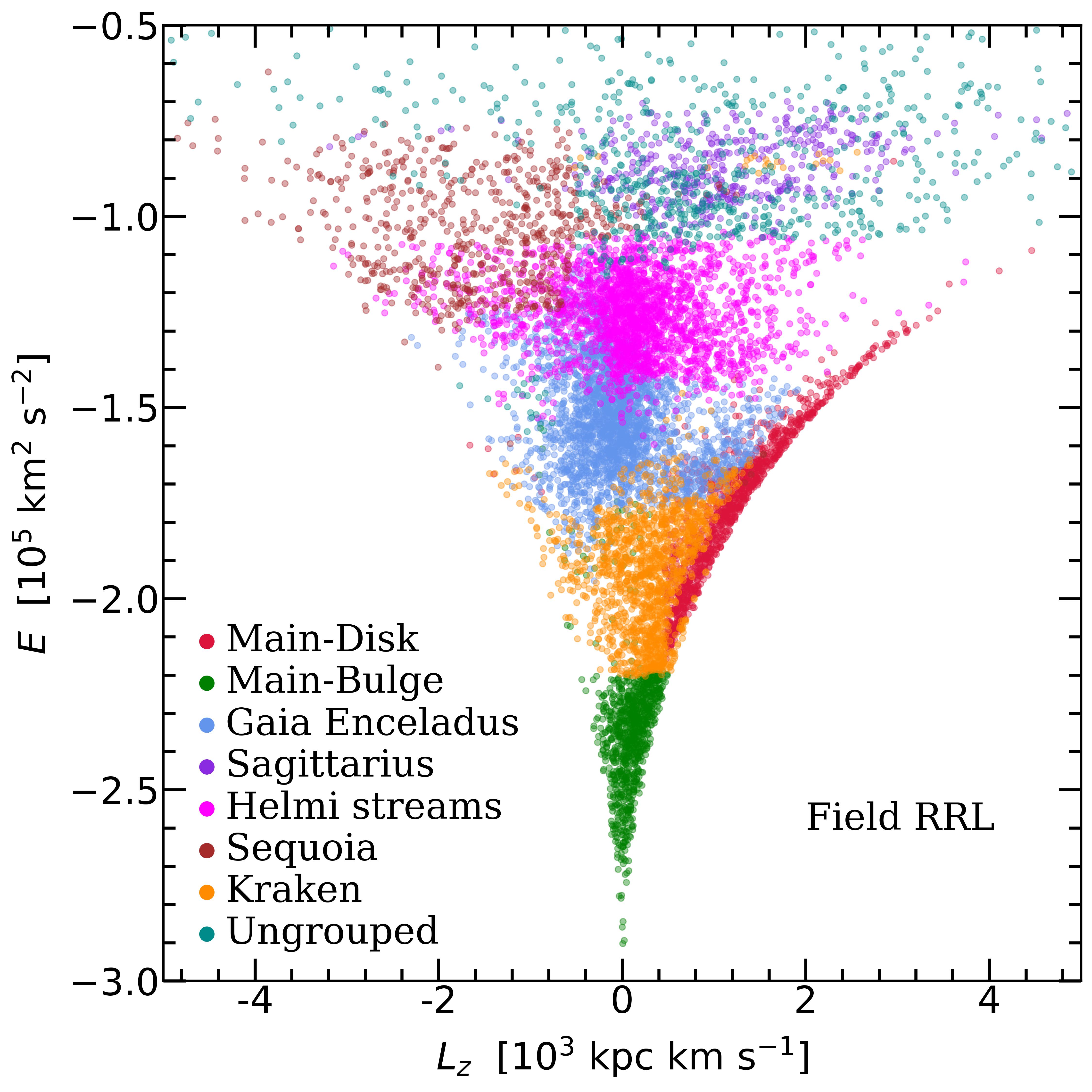}}
    }
    \caption{\textit{Top left} ($E$, $L_z$) diagram for the GGCs considered in this work. OoI, OoII and GGCs without Oosterhoff type are shown with filled circles,  diamonds and crosses, respectively. The GGCs are classified in terms of \textit{In-Situ/Accreted} according to \citet{belokurov2024situ} and plotted in different colours (see labels).
    \textit{Bottom left} as before but for the classification in eight \textit{Progenitors} by \citet{callingham2022chemo}. Each \textit{Progenitor} is identified with a different colour in the Figure (see labels). 
    \textit{Top right} as in the \textit{Top left} but for the field RRLs;  
    \textit{Bottom right} as in the \textit{Bottom left} but for the field RRLs.
        \label{Fig:e-lz-class-GGC-RRL}}
\end{figure*}

\begin{figure*}[htbp]
    \centering
    \vbox{
    \hbox{
    \includegraphics[width=0.5\textwidth, height=0.5\textheight, keepaspectratio]{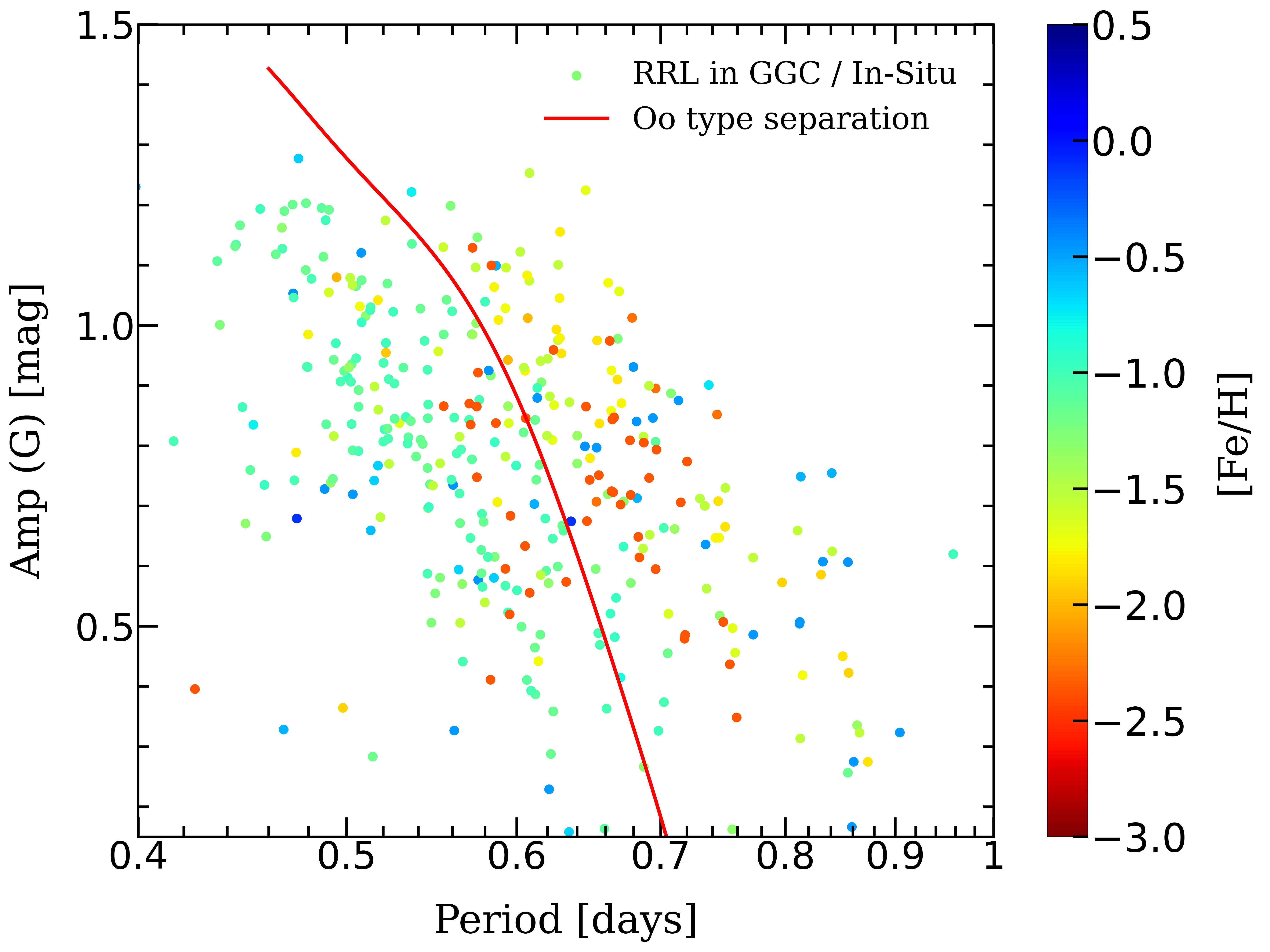}
    \includegraphics[width=0.5\textwidth, height=0.5\textheight, keepaspectratio]{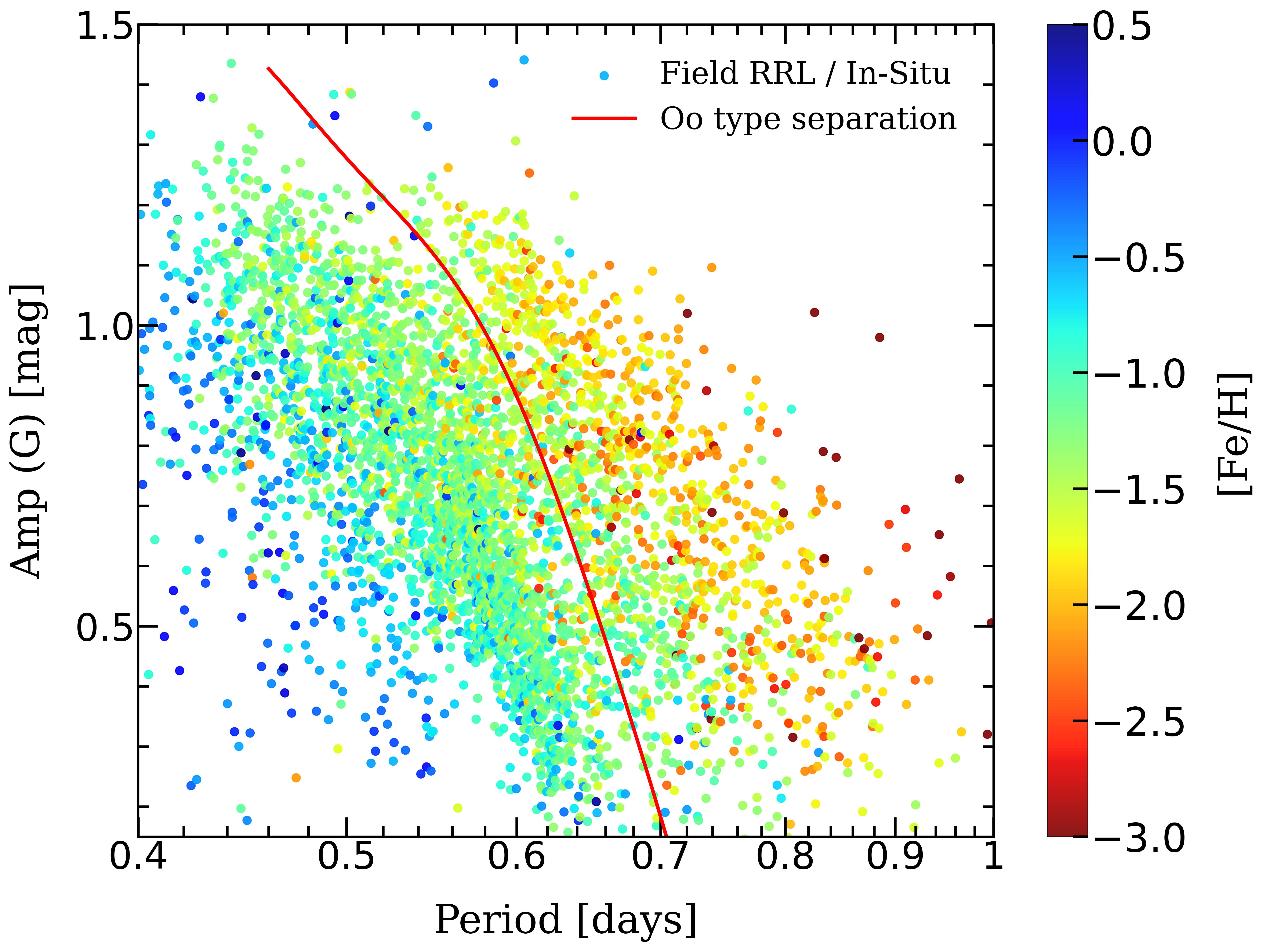}}
    \hbox{
    \includegraphics[width=0.5\textwidth, height=0.5\textheight, keepaspectratio]{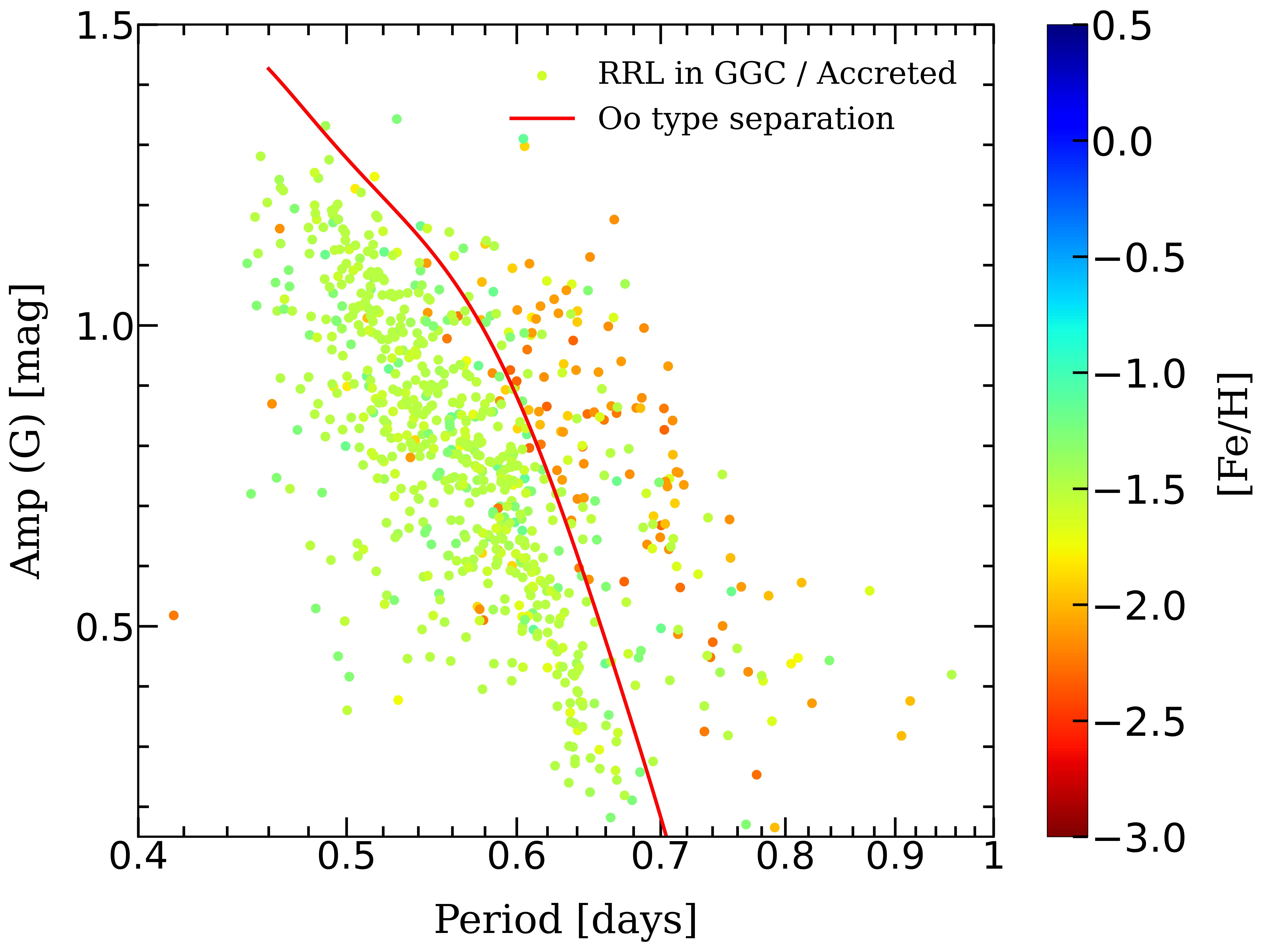}
    \includegraphics[width=0.5\textwidth, height=0.5\textheight, keepaspectratio]{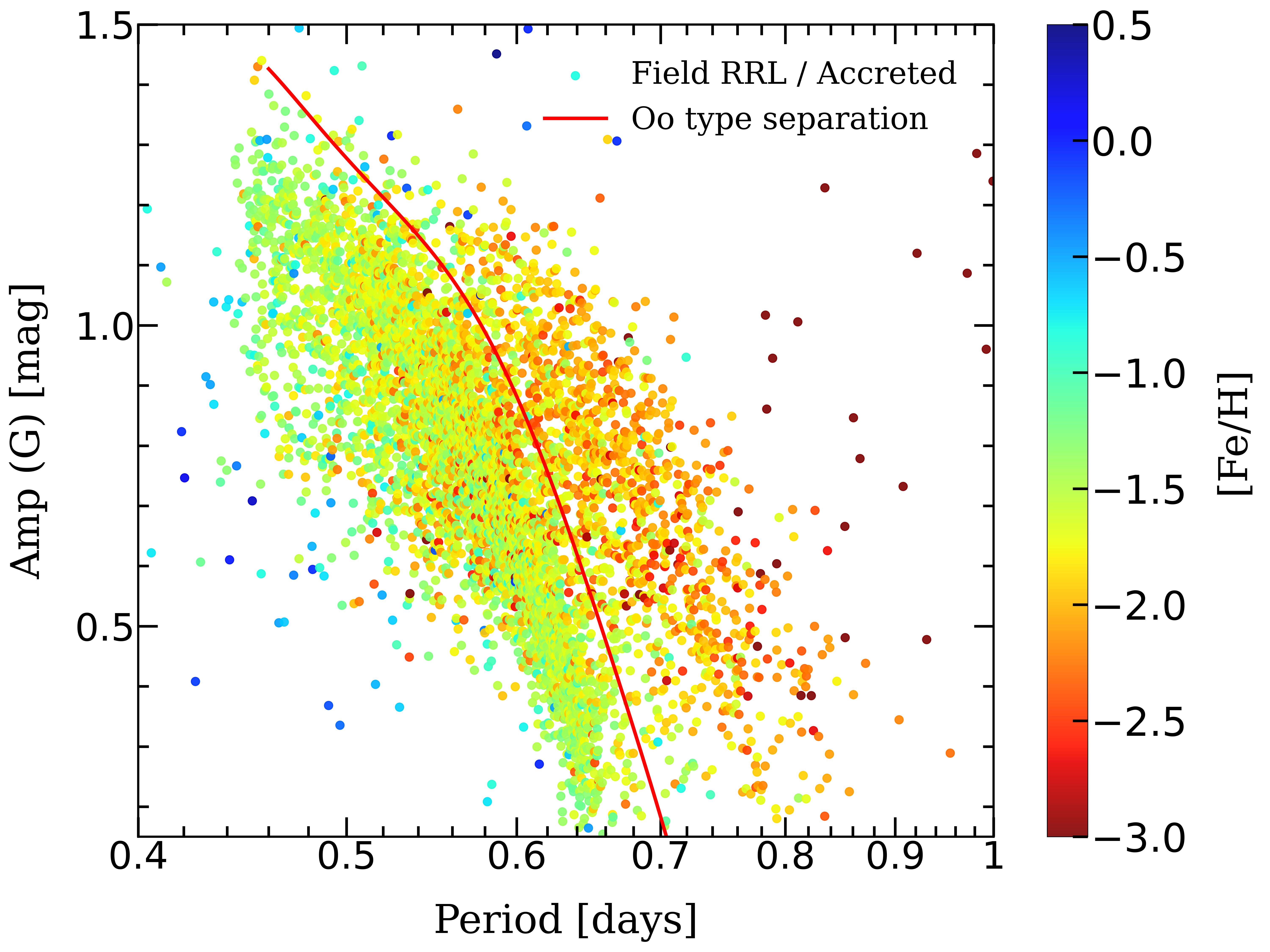}}
    }
    \caption{
    \textit{Top left} Bailey diagram of RRLs in GGCs associated with the \textit{In-Situ} population.
    \textit{Bottom left} Bailey diagram of RRLs in GGCs associated with the \textit{Accreted} population.
    \textit{Top right} Bailey diagram of field RRLs associated with the \textit{In-Situ} population.
    \textit{Bottom right} Bailey diagram of field RRLs associated with the \textit{Accreted} population. Sources are colour-coded according to their metallicity. The separation red curve between the two types of Oosterhoff dichotomy is described by the following polynomial: $y = -8.70 \cdot 10^5\,x^7 - 1.63 \cdot 10^6\,x^6 -1.28\cdot10^6\,x^5 -5.44\cdot10^5\,x^4 -1.35\cdot10^5\, x^3 -1.95\cdot10^4\,x^2-1.54\cdot10^3\,x -52.14$.}
        \label{Fig:Bailey_GGC_field_RRL}
\end{figure*}

\section{Field RRL sample}
\label{Sec:Field_RRLs_sample}

In addition to the GGC sample, we also considered field RRLs. Collecting the six parameters needed to study the dynamics of these objects is not as straightforward as for the GGCs because there is no database listing all together the necessary information.
We can use $Gaia$ to find positions and proper motions of all RRLs in our study, but the RVs are only available for a fraction of them in the $Gaia$ DR3 (Table~\ref{tab:radial_velocity}). Furthermore, given the large distances at which many of our targets are located, we found that often the $Gaia$ parallaxes were not informative, leading to unreliable individual distance measurements. Therefore, we decided to use the property of RRLs as excellent standard candles to calculate their distances using the luminosity-metallicity relation that RRLs conform to.
As shown in the next section, to infer distances from this relation we need individual metal abundances, that can be obtained from the $\phi_{31}$ and $R_{21}$ parameters of the light curve Fourier decomposition. Therefore for the selection of field RRLs, we first considered the $Gaia$ DR3 sample of RRLs resulting from the Specific Object Study (SOS) pipeline \citep[{\tt vari\_rrlyrae} table in the $Gaia$ DR3 archive][]{Clementini2023} and consisting of about 270,000 objects. We did not take into account other photometric surveys to keep the homogeneity of our photometric sample.
Among the $Gaia$ RRLs, we selected fundamental mode pulsators (RRab) because they "much neatly" show the Oostherhoff dichotomy in the Bailey diagram \citep[e.g. Fig.35 in ][]{Clementini2023} and retained only DR3 RRLs with $P>0.4$ days to avoid possibly misclassified first overtone pulsators. In addition, we retained only those RRab for which Fourier parameters are available in the DR3 {\tt vari\_rrlyrae} table. The third selection was based on the availability of the average RV in the $Gaia$ DR3 archive which occurs for only a few thousands fundamental mode RRLs. To enlarge the sample, we therefore scanned the literature and gathered additional RRLs having measured RV (see Sect.~\ref{Sec:Sigle_Epoch_Radial_Velocities} for details). Our final sample of field RRLs includes 10,138 variable stars. Their identifications, positions, metallicitiy, Fourier amplitudes and other parameters are listed in Table~\ref{Tab:5field_RRLs_dataset}.

\subsection{Determination of distances to the RRLs}
\label{Sec:Determination_of_the_distances_for_RRLs}

To determine the distances to the field RRLs, we adopted the luminosity-metallicity in the G band relation ($M_G$--[Fe/H]) from \citet{li2023photometric}. This relation can provide distances to the investigated RRLs from their dereddened apparent magnitudes and metallicity. G band absorption values for the RRLs were taken from the {\tt vari\_rrlyrae} table in the $Gaia$ DR3 archive \citep[][]{Clementini2023}.
The same $Gaia$ catalogue also provides individual metallicity estimates for the RRLs. However, the Fourier parameter-metallicity calibration used in $Gaia$ \citep[based on the work by][]{nemec2013metal} overestimates the metallicity for [Fe/H] values larger than $\sim -1.3 $ dex \citep[see sect 5.1 of][]{Clementini2023}. We calculated the metallicities using the equation from \citet{li2023photometric} valid for RRab stars, connecting the [Fe/H] value to the period $P$ and the Fourier parameters $R_{21}$ and $\phi_{31}$. 
The individual RRL distances are then evaluated from the inferred metal abundances and the $Gaia$ de-reddened $G$ magnitudes.
The errors associated with distance determinations are evaluated by propagating uncertainties in the evaluated $M_G$ and the dereddened apparent magnitude ($G_0$), through the distance modulus relation. The first term is the RMS of the adopted $M_G$-$[Fe/H]$ relation, taken from \citet{li2023photometric}, while the latter is calculated by propagating the errors on the two quantities measured by $Gaia$ ($G$ and $A(G)$).
All the inferred individual distances are reported in Table~\ref{Tab:5field_RRLs_dataset}.

\subsection{Random Forest to classify field RRLs}
\label{Sec:Random_Forest_to_classify_field_RRLs}

The last piece of information missing to procede using the field RRLs for our purposes is their classification in terms of \textit{In-Situ} or \textit{Accreted} populations, similar to what is available in the literature for GGCs.   
To this aim, we adopted the same two classifications used for GGCs, according to \citet{belokurov2024situ} and \citet{callingham2022chemo}, respectively, on the assumption that the satellite galaxies which were involved in merging episodes with the MW carried with them both GGCs and field RRLs, which thus would occupy the same region in the IoM space. We adopted the classifications used for GGCs as training sets to identify the different regions in IoM space where the field RRLs would be located using the {\tt scikit-learn} library of the {\tt RandomForestClassifier} \citep[][]{breiman2001random}, with the hyper-parameters reported in \ref{tab:hyperparameters}.
In particular, we trained the model using GGCs data and the attributes ($E$, $J_r$, $J_{\phi}$, $J_z$) calculated with {\tt Galpy} based on the B\&Co database. According to the classification scheme by \citep{belokurov2024situ}, we assigned the labels \textit{In-Situ} and \textit{Accreted}. Then following \citet{callingham2022chemo}, we assigned the eight progenitors labelled by these Authors (see Table~\ref{tab:prog}). The test results are reported in Table\,\ref{tab:in-situ-accreted-randomforest} and Table\,\ref{tab:prog-randomforest}, based on \textit{In-Situ/Accreted} and \textit{Progenitors} classifications, respectively. 
As a result of this procedure, the top left panel of Fig.~\ref{Fig:e-lz-class-GGC-RRL} shows the distribution of GGCs in the ($E$, $L_z$) plane used as a training set to classify the field RRLs. The objects formed in the pristine Galaxy and those carried into the MW by merging satellites are labelled as \textit{In-Situ} and \textit{Accreted}, respectively. The Figure also shows for completeness the Oosterhoff classification by \citet{van2011some,stobie1971difference}. 
Similarly, in the bottom left panel of Fig.~\ref{Fig:e-lz-class-GGC-RRL} 
we show the GGCs in the same plane as above but with a different colour code, namely according to the classification in eight progenitors by \citet{callingham2022chemo} (see Table~\ref{tab:prog}). 
The classification of the field RRLs based on the GGCs is shown in the right top and bottom panels of Fig.~\ref{Fig:e-lz-class-GGC-RRL} for the two classification methods \textit{In-Situ/Accreted} and \textit{Progenitors}, respectively. The field RRLs are colour-coded according to the classification obtained with the {\tt RandomForestClassifier}, using the same colours as for the GGCs in the left panels. The percentage of field RRLs and GGCs classified as \textit{In-Situ/Accreted} is shown in  Fig.~\ref{Fig:in_situ_accreted_bar_chart}.

\section{Is the Oosterhoff dichotomy linked to the past merging history of the MW?}
\label{Sec:Is_the_Oosterhoff_dichotomy_linked_to_the_past_merging_history_of_the_Milky_Way?}

Once classified the field RRLs in terms of {\it In-situ/Accreted} populations, we can look for a possible connection with the Oosterhoff dichotomy. 
To assign the Oosterhoff type we use the Bailey diagram as shown in the top and bottom panels of Fig.~\ref{Fig:Bailey_GGC_field_RRL} for the \textit{In-Situ} and \textit{Accreted} populations, respectively. RRLs are colour-coded according to their metallicity. 
The two panels show remarkably different distributions. The \textit{In-Situ} population shows a wide and continuous range of metallicities from low-[Fe/H] (at longer periods) to high-[Fe/H] (at shorter periods), with no sign of Oosterhoff dichotomy, as the distribution of pulsators is almost uniform at all periods. The bottom panel seems to tell a different story for the RRLs classified as \textit{Accreted}. Indeed, they show different Oosterhoff types, together with a sharp separation in metallicity, with the OoI and OoII showing intermediate ([Fe/H]$\sim-1.5$ dex) and metal-poor ([Fe/H]$>-2.0$ dex) metallicity, respectively. Also, the total metallicity range spanned by the \textit{Accreted} RRLs is much reduced compared with the \textit{In-Situ} population. 
A further confirmation of this occurrence is shown in Fig.~\ref{Fig:delta_log_p}, where we have represented the Bailey diagram as a histogram. Again, we find two sharper peaks for the \textit{Accreted} population and a smoother distribution for that \textit{In-Situ}.
These results suggest that the Oosterhoff dichotomy was imported into the MW by the merging events that shaped the Galaxy.\\
We then checked if the same trend was present also among the GGC RRLs. We thus cross-matched the $Gaia$ RRLs DR3 catalogue with the database of RRLs in GGCs by \citet{clement2001variable}, finding 1157 matches for the fundamental mode RRLs, and gathered the GGCs metallicities from the latest version (December 2010) of the \citet{Harris1996} database. The adopted data for the RRLs in GGCs are shown in Table~\ref{Tab:5RRLs_in_GGCs_dataset}. The classification of the GGCs in terms of \textit{In-Situ/Accreted} is the same as in the previous sections (see Table~\ref{Tab:5GGCs_dataset}). Similarly to the field RRLs, the results are shown in the left panels of Fig.~\ref{Fig:Bailey_GGC_field_RRL}. Even if statistics is not large, it looks like also for GGCs the Oosterhoff dichotomy was imported into the MW by ancient merging events. 

We repeated this exercise for each of the eight progenitors from \citet{callingham2022chemo} to which we have associated our RRLs sample. The Bailey diagrams for the three progenitors of Galactic origin (equivalent to the \textit{In-Situ} population) are shown in Fig.~\ref{Fig:Bailey_field_RRL_prog_in_situ}, while those associated with the five progenitors of extragalactic origin (equivalent to the \textit{Accreted} population) are shown in Fig.~\ref{Fig:Bailey_field_RRL_prog_acc}. Again, we can see that the "Galactic" progenitors do not show the Oosterhoff dichotomy, which is instead present in each of the five "extragalactic" progenitors. \\
This further confirms that there may be a strict link between the past merging history of the MW and the Oosterhoff dichotomy. 
Note that our result is not in contradiction with findings by \citet{Fabrizio2019,Fabrizio2021}. Indeed, we confirm their results if we consider the {\it In-Situ} RRL populations only. 

However, our results now raise new questions. In first place: why the progenitors that merged in the MW long ago did show the Oosherhoff dichotomy and the dwarf spheroidal galaxies that orbit the Galaxy today do not? And, how would OoIII GGCs like NGC6388 and NGC6441 fit into such a scenario?

The answer to these questions is beyond the scope of the present letter. However, we note that among the many ultra-faint dwarf galaxies (UFDs) identified within the halo of the MW (and M31) in the last 20 years those more dispersed and poorly populated seem to host RRLs conforming to an Oosterhoff dichotomy, hence somehow supporting our interpretation of the dichotomy as an "alien" imported from the outside. Luckily, the improved data that will be released by $Gaia$ DR4, along with next large and deep photometric surveys such as the Legacy Survey of Space and Time (LSST) at the Vera Rubin telescope, and big spectroscopic surveys such as those foreseen with WEAVE (WHT Enhanced Area Velocity Explorer)\footnote{https://www.ing.iac.es/astronomy/instruments/weave/weaveinst.html} and 4MOST (4-metre Multi-Object Spectroscopic Telescope)\footnote{https://www.eso.org/sci/facilities/develop/instruments/4MOST.html}, in a couple of years will allow us to calculate the IoM for tens of thousands of RRLs, thus providing new crucial insights in the formation and evolution of the Galactic halo.

\begin{acknowledgements}We warmly thank our anonymous Referee for their suggestions.
This research has made use of the
SIMBAD database operated at CDS, Strasbourg, France.
We acknowledge support from Project PRIN MUR 2022 (code 2022ARWP9C) “Early Formation and Evolution of Bulge and HalO (EFEBHO)", PI: Marconi, M.,  funded by European Union – Next Generation EU; INAF Large grant 2023 MOVIE (PI: M. Marconi); INAF GO-GTO grant 2023 “C-MetaLL - Cepheid metallicity in the Leavitt law” (P.I. V. Ripepi).

This work has made use of data from the European Space
Agency (ESA) mission Gaia (https://www.cosmos.esa.int/gaia),
processed by the Gaia Data Processing and Analysis Consortium (DPAC,
https://www.cosmos.esa.int/web/gaia/dpac/consortium). Funding
for the DPAC has been provided by national institutions, in particular, the
institutions participating in the Gaia Multilateral Agreement.
This research was supported by the Munich Institute for Astro-, Particle and BioPhysics (MIAPbP), which is funded by the Deutsche Forschungsgemeinschaft (DFG, German Research Foundation) under Germany´s Excellence Strategy – EXC-2094 – 390783311. This work is supported through an internship programme at the European Southern Observatory (ESO) in Garching, Germany.

\end{acknowledgements}


%
%

\bibliographystyle{aa}
\bibliography{paperOosterhoff.bib}


\begin{appendix}

\section{Radial Velocities}
\label{Sec:Sigle_Epoch_Radial_Velocities}
Average RV measurements from the $Gaia$ catalogue are available for only 3442 fundamental-mode pulsators (Table~\ref{tab:radial_velocity}). Therefore to complement our sample we considered several large spectroscopic surveys namely, APOGEE, SDSS, LAMOST LRS and GALAH (see Table~\ref{tab:radial_velocity} for details). We performed the following steps: i) we cross-matched the catalogue of RRLs selected as explained above with each of these surveys; ii) in case of multiple matches for the same star, we retained only one measurement, using preferentially the data in the order from top to bottom of Table~\ref{tab:radial_velocity}. The same table lists the number of RRLs with RV data from each survey.  

We estimate the RMSE (Root-Mean-Square-Error) value for the RV data of each survey using Gaia data as the model:
\begin{equation}
    RMSE_x = \sqrt{\sum_{i=1}^{N}\frac{{(RV_{x,i} - RV_{\textit{Gaia},i})}^2}{N}}
\end{equation}
where \textit{x} is referred to a generic survey excluded \textit{Gaia} and \textit{N} is the number of stars which have RV data from both survey \textit{x} and \textit{Gaia}.
The ratio between the RMSE and the error of each star are major then one, so this leads us to consider the RMSE calculated for the survey as the error related to the RV of each star in that survey.
All the adopted RV measurements and relative errors are listed in Table~\ref{Tab:5field_RRLs_dataset}.

\section{Tables and figures accompanying this letter}

\begin{table*} 
    \footnotesize\setlength{\tabcolsep}{5pt} 
    \caption{Number of stars coming from each survey.} 
    \label{tab:radial_velocity}
        \begin{tabular}{lrrl} 
            \hline  
            \multicolumn{1}{c}{Survey} & \multicolumn{1}{c}{N RRab}\\
            \hline
            $Gaia$ DR3 & 3442\\
            APOGEE & 1921\\ 
            SDSS & 2458\\
            LAMOST LRS & 2308\\
            GALAH & 9\\
            Total & 10,138\\
            \hline
        \end{tabular}
    \tablefoot{Each survey is listed with its acronym as follows: $Gaia$ DR3 {\href{https://www.cosmos.esa.int/web/gaia/dr3}{https://www.cosmos.esa.int/web/gaia/dr3}}; Apache Point Observatory Galactic Evolution Experiment (APOGEE) {\href{https://www.sdss4.org/dr17/irspec/}{https://www.sdss4.org/dr17/irspec/}}; Sloan Digital Sky Survey (SDSS) {\href{https://www.sdss.org/dr18/}{https://www.sdss.org/dr18/}}; Large Sky Area Multi-Object Fiber Spectroscopic Telescope Low Resolution Spectrograph (LAMOST LRS) {\href{https://www.lamost.org/dr8/v2.0/doc/lr-data-production-description}{https://www.lamost.org/dr8/v2.0/doc/lr-data-production-description}}; Galactic Archaeology with HERMES (GALAH) {\href{https://www.galah-survey.org/dr3/overview/}{https://www.galah-survey.org/dr3/overview/}}.} 
\end{table*}

\begin{table*} 
    \footnotesize\setlength{\tabcolsep}{5pt} 
    \caption{Hyperparameters used in {\tt RandomForestClassifier}.} 
    \label{tab:hyperparameters}
    \begin{tabular}{lrl} 
        \hline  
        \multicolumn{1}{c}{Hyperparameter} & \multicolumn{1}{c}{Value} \\
        \hline
        \texttt{n\_estimators} & 100\\
        \texttt{criterion} & gini\\
        \texttt{max\_depth} & None\\
        \texttt{min\_samples\_split} & 2\\
        \texttt{min\_samples\_leaf} & 1\\
        \texttt{min\_weight\_fraction\_leaf} & 0.0\\
        \texttt{max\_features} & auto\\
        \texttt{max\_leaf\_nodes} & None\\
        \texttt{min\_impurity\_decrease} & 0.0\\
        \texttt{bootstrap} & True\\
        \texttt{oob\_score} & False\\
        \texttt{n\_jobs} & None\\
        \texttt{random\_state} & 42\\
        \texttt{verbose} & 0\\
        \texttt{warm\_start} & False\\
        \texttt{class\_weight} & None\\
        \hline
    \end{tabular}
    \tablefoot{Hyperparameters are listed with their default values as used in the {\tt RandomForestClassifier} from scikit-learn.} 
\end{table*}

\begin{table*}\footnotesize\setlength{\tabcolsep}{5pt} 
    \caption{Classification report of field RRLs for the method by \citet{belokurov2024situ}.}
    \label{tab:in-situ-accreted-randomforest}
    \begin{tabular}{cccc}
        \hline
        \multicolumn{1}{c}{Class} & \multicolumn{1}{c}{precision} & \multicolumn{1}{c}{recall} & \multicolumn{1}{c}{f1-score}\\   
        \hline
           0    &   1.00   &   1.00  &    1.00\\
           1    &   1.00   &   1.00  &    1.00\\
        \hline
        accuracy    &   1.00\\
        \hline
    \end{tabular}
    \tablefoot{The results from the test phase of the model trained with GGCs labeled according to the classification method by \citet{belokurov2024situ} indicate two distinct populations. The label "0" corresponds to the \textit{Accreted} population, while the label "1" corresponds to the \textit{In-Situ} population. Consider the following abbreviations: TP (True-Positive), FP (False-Positive), TN (True-Negative), and FN (False-Negative).
    We also recall the following definitions: $\text{precision} = \frac{TP}{TP\,+\,FP}$ , $\text{recall} = \frac{TP}{TP\,+\,FN}$ , $\text{f1- score} = \frac{2(precision \cdot recall)}{precision \,+\,recall}$, $\text{accuracy} = \frac{TP\,+\,TN}{TP\,+\,TN\,+\,FP\,+\,FN}$}
\end{table*}

\begin{table*}\footnotesize\setlength{\tabcolsep}{5pt} 
    \caption{Classification report of field RRLs for the method by \citet{callingham2022chemo}.}
    \label{tab:prog-randomforest}
        \begin{tabular}{cccc}
        \hline
        \multicolumn{1}{c}{Class} & \multicolumn{1}{c}{precision}  &  \multicolumn{1}{c}{recall} & \multicolumn{1}{c}{f1-score}\\   
        \hline
         G-E    &   1.00    &  1.00   &   1.00 \\        
         H-E    &   0.50    &  0.50   &   0.50 \\        
         H99    &   0.50    &  1.00   &   0.67 \\        
         L-E    &   0.75    &  0.75   &   0.75 \\        
         M-B    &   1.00    &  0.82   &   0.90 \\       
         M-D    &   0.50    &  1.00   &   0.67 \\  
         Seq    &   1.00    &  0.33   &   0.50 \\        
        \hline
        accuracy    &   0.78 \\
        \hline
    \end{tabular}
    \tablefoot{The results from the test phase of the model trained with GGCs labeled according to the classification method by \citet{callingham2022chemo}. Abbreviations refer to the \textit{Progenitors} list in Table\,\ref{tab:prog}. See \ref{tab:in-situ-accreted-randomforest} for the definition of \textit{precision}, \textit{recall}, \textit{f1-score} and \textit{accuracy} }
\end{table*}

\begin{table*} \footnotesize\setlength{\tabcolsep}{5pt} 
    \caption{List of the 8 \textit{Progenitors} identified by \citet{callingham2022chemo} with the respective acronyms.} 
    \label{tab:prog}
        \begin{tabular}{lrcl} 
        \hline  
        \multicolumn{1}{c}{Progenitors}\\
        \hline
        Main-Disk & (M-D)\\
        Main-Bulge & (M-B)\\
        Gaia-Enceladus & (G-E)\\
        Sagittarius & (Sag)\\
        Helmi Streams & (H99)\\
        Sequoia & (Seq)\\
        Kraken/Low-Energy & (L-E)\\
        Ungrouped/High Energy & (H-E)\\
        \hline
    \end{tabular}
\end{table*}

\begin{table*} \footnotesize\setlength{\tabcolsep}{5pt} 
    \caption{List of considered GGCs.} 
    \label{Tab:5GGCs_dataset}
        \begin{tabular}{lccrrrrrrrrrl} 
        \hline  
        \multicolumn{1}{c}{Name} & \multicolumn{1}{c}{In-situ/Accr.} & \multicolumn{1}{c}{Oo Type} & \multicolumn{1}{c}{Prog.} & \multicolumn{1}{c}{$N_{RRL}$} & \multicolumn{1}{c}{RA} & \multicolumn{1}{c}{DEC} & \multicolumn{1}{c}{$d$} & \multicolumn{1}{c}{$\mu_{\rm RA}$} & \multicolumn{1}{c}{$\mu_{\rm DEC}$} & \multicolumn{1}{c}{RV}\\
         \multicolumn{1}{c}{} & \multicolumn{1}{c}{} & \multicolumn{1}{c}{} & \multicolumn{1}{c}{} & \multicolumn{1}{c}{} & \multicolumn{1}{c}{[deg]} & \multicolumn{1}{c}{[deg]} & \multicolumn{1}{c}{[kpc]} & \multicolumn{1}{c}{[mas/yr]} & \multicolumn{1}{c}{[mas/yr]} & \multicolumn{1}{c}{[km/s]}\\
        \hline
            NGC104 & 1 & 0 & M-D & 48 & 6.024 & -72.081 & 4.52 $\pm$ 0.03 & 5.252 $\pm$ 0.021 & -2.551 $\pm$ 0.021 & -17.44 $\pm$ 0.16 \\
            NGC1261	& 0 & 1 & G-E & 9 & 48.068 & -55.216 & 16.40 $\pm$ 0.19 & 1.596 $\pm$ 0.025 & -2.064 $\pm$ 0.025 & 71.34 $\pm$ 0.21 \\
            NGC1851	& 0 & 1 & G-E & 21 & 78.528 & -40.047 & 11.95 $\pm$ 0.13 & 2.145 $\pm$ 0.024 & -0.65 $\pm$ 0.024 & 321.39 $\pm$ 1.55 \\
            NGC1904	& 0 & 2 & H99 & 5 & 81.044 & -24.524 & 13.08 $\pm$ 0.18 & 2.469 $\pm$ 0.025 & -1.594 $\pm$ 0.025 & 205.75 $\pm$ 0.2 \\
            NGC2298 & 0 & 0 & G-E & 9 & 102.248 & -36.005 & 9.83 $\pm$ 0.17 & 3.32 $\pm$ 0.025 & -2.175 $\pm$ 0.026 & 147.15 $\pm$ 0.57 \\
        \hline
    \end{tabular}
\tablefoot{Name of the GGC; classification of GGCs according to \citet{belokurov2024situ}: \textit{In-Situ}=1 \textit{Accreted}=0; classification of GGCs according to \citet{callingham2022chemo}, meaning of the acronyms is provided in Table~\ref{tab:prog}; Oo type lists the classification of GGCs in Oosterhoff types according to \citet{van2011some} and \citet{stobie1971difference}: 0, 1 and 2 refer to no-classification, OoI and OoII, respectively; $N_{RRL}$ is the number of RRLs for each GGCs collected from \citet{clement2001variable}; RA, DEC, $d$, $\mu_{\rm RA}$, $\mu_{\rm DEC}$ and RV are equatorial coordinates (J2000), distances, proper motions and radial velocities of the GGCs with relative errors according to the B\&Co catalogue. Note that the GGC Mercer5 was not classified by \citet{belokurov2024situ}. We classified this GGC as \textit{In-Situ} based on its position in the ($E$, $L_z$) diagram. \\
Only a portion of the table is shown here to guide the reader about its content. A machine-readable version of the full table will be published at the CDS (Centre de Données astronomiques de Strasbourg, https://cds.u-strasbg.fr/)
} 
\end{table*}

\begin{table*}[ht]
    \footnotesize\setlength{\tabcolsep}{2pt} 
    \caption{Data of the field RRLs used in this letter.} 
    \label{Tab:5field_RRLs_dataset}
        \begin{tabular}{cccccccccccccccccccccccc} 
        \hline  
        \multicolumn{1}{c}{source\_id} & \multicolumn{1}{c}{RA} & \multicolumn{1}{c}{DEC} & \multicolumn{1}{c}{$d$} & \multicolumn{1}{c}{$\mu_{\rm RA}$} & \multicolumn{1}{c}{$\mu_{\rm DEC}$} & \multicolumn{1}{c}{RV} & \multicolumn{1}{c}{G} & \multicolumn{1}{c}{[Fe/H]} & \multicolumn{1}{c}{M$_G$} & \multicolumn{1}{c}{...}\\
        \multicolumn{1}{c}{} & \multicolumn{1}{c}{[deg]} & \multicolumn{1}{c}{[deg]} & \multicolumn{1}{c}{[kpc]} & \multicolumn{1}{c}{[mas/yr]} & \multicolumn{1}{c}{[mas/yr]} & \multicolumn{1}{c}{[km/s]} & \multicolumn{1}{c}{[mag]} & \multicolumn{1}{c}{dex} & \multicolumn{1}{c}{[mag]} & \multicolumn{1}{c}{}\\
        \hline
            2098934029183133696	& 287.7098 & 37.4616 & 3.593$\pm$0.217 & 0.135$\pm$0.012 & 9.173$\pm$0.014 & $-$577.9$\pm$3.5 & 13.556$\pm$0.007 & $-$1.79 & 0.480 & ...\\
            2706810985485718400 & 335.3695 & 3.1660 & 5.672$\pm$0.547 & $-$1.308$\pm$0.030 & $-$7.744$\pm$0.032 & $-$564.7$\pm$13.1 & 14.625$\pm$0.014 & $-$2.00& 0.405 & ...\\
            2301463413882932992 & 297.2065 & 82.2221 & 3.803$\pm$0.263 & $-$7.359$\pm$0.015 & 5.243$\pm$0.015 & 495.9$\pm$3.6 & 13.727$\pm$0.013 & $-$1.47& 0.592 & ...\\
            563515896970574208	& 20.1593 & 77.0142 & 3.044$\pm$0.188 & $-$1.748$\pm$0.023 & 6.612$\pm$0.025 & $-$448.5$\pm$5.5 & 15.120$\pm$0.005 & $-$1.24& 0.6705 & ...\\
            1345084633559363072 & 269.5150 & 41.8304 & 4.020$\pm$0.243 & $-$3.291$\pm$0.013 & $-$16.372$\pm$0.014 & $-$446.1$\pm$3.2 & 13.519$\pm$0.006 & $-$1.76& 0.488& ... \\
        \hline
        \hline
        ...& \multicolumn{1}{c}{Period} & \multicolumn{1}{c}{R$_{21}$} & \multicolumn{1}{c}{R$_{31}$} & \multicolumn{1}{c}{$\phi_{21}$} & \multicolumn{1}{c}{$\phi_{31}$} & \multicolumn{1}{c}{Amp(G)} & \multicolumn{1}{c}{A(G)} & \multicolumn{1}{c}{In-Situ/} & \multicolumn{1}{c}{Prog.} & \multicolumn{1}{c}{Alt.} \\
         & \multicolumn{1}{c}{[days]} & \multicolumn{1}{c}{} & \multicolumn{1}{c}{} & \multicolumn{1}{c}{} & \multicolumn{1}{c}{} & \multicolumn{1}{c}{[mag]} & \multicolumn{1}{c}{[mag]} & \multicolumn{1}{c}{Acc.} & \multicolumn{1}{c}{} & \multicolumn{1}{c}{Prog.} \\
        \hline
            ... & 0.700 & 0.390$\pm$0.015 & 0.170$\pm$0.014 & 4.299$\pm$0.023 & 2.683$\pm$0.090 & 0.392$\pm$0.009 & 0.274$\pm$0.053 & 0 & Seq & H-E\\
            ... & 0.685 & 0.426$\pm$0.078 & 0.283$\pm$0.084 & 4.114$\pm$0.130 & 2.367$\pm$0.574 & 0.571$\pm$0.053 & 0.416$\pm$0.171 & 0 & Seq & H-E\\
            ... & 0.506 & 0.440$\pm$0.071 & 0.312$\pm$0.071 & 4.128$\pm$0.244 & 1.897$\pm$0.286 & 0.890$\pm$0.061 & 0.216$\pm$0.090 & 0 & Seq & G-E\\
            ... & 0.635 & 0.354$\pm$0.033 & 0.164$\pm$0.027 & 4.313$\pm$0.077 & 2.872$\pm$0.184 & 0.295$\pm$0.025 & 1.872$\pm$0.060 & 0 & G-E & H99\\
            ... & 0.692 & 0.346$\pm$0.022 & 0.130$\pm$0.023 & 4.307$\pm$0.104 & 2.704$\pm$0.405 & 0.329$\pm$0.020 & 0.009$\pm$0.053 & 0 & H99 & Seq\\
        \hline
    \end{tabular}
\tablefoot{source\_id is the $Gaia$ identifier of the variable; RA and DEC are the equatorial coordinates (J2000); $d$ is the distance calculated as described in Sect.~\ref{Sec:Determination_of_the_distances_for_RRLs}; $\mu_{\rm RA}$ and $\mu_{\rm DEC}$ are the $Gaia$ proper motions; RV lists the radial velocities collected as reported in Sect.~\ref{Sec:Sigle_Epoch_Radial_Velocities}; $G$, $A(G)$, Period, $R_{21}$, $R_{31}$, $\phi_{21}$, $\phi_{31}$ and Amp(G) are the magnitudes in the $G$ band, the extinction in the same band, the periods, the Fourier parameters and the amplitudes in the $G$ band according to $Gaia$ DR3 \citep[][]{Clementini2023}; [Fe/H] is the metallicity derived in Sect.~\ref{Sec:Determination_of_the_distances_for_RRLs}; $M_G$ is the absolute magnitude in the $G$ band; Prog. is the most probable class and Alt. Prog. the secondary probable class assigned in terms of \citet{callingham2022chemo} progenitors obtained in Sect.\ref{Sec:Random_Forest_to_classify_field_RRLs}; \textit{In-Situ/Acc.} provides the classification in terms of \textit{In-Situ/Accreted} populations by \citet{belokurov2024situ}, as determined in Sect.\ref{Sec:Random_Forest_to_classify_field_RRLs}(\textit{In-Situ}=1; \textit{Accreted}=0).
Only a portion of the table is listed here to guide the reader about its content. A machine-readable version of the full table will be published at the CDS (Centre de Données astronomiques de Strasbourg, https://cds.u-strasbg.fr/)} 
\end{table*}

\begin{table*} 
    \footnotesize\setlength{\tabcolsep}{5pt} 
    \caption{Data for the RRLs in GGCs.} 
    \label{Tab:5RRLs_in_GGCs_dataset}
        \begin{tabular}{lccrrrrrrrl} 
        \hline  
        \multicolumn{1}{c}{Name} & \multicolumn{1}{c}{Var} & \multicolumn{1}{c}{In-situ/Acc.} & \multicolumn{1}{c}{Prog.} & \multicolumn{1}{c}{RA} & \multicolumn{1}{c}{DEC} & \multicolumn{1}{c}{Period} & \multicolumn{1}{c}{Amp(G)} & \multicolumn{1}{c}{[Fe/H]}\\
         \multicolumn{1}{c}{} & \multicolumn{1}{c}{} & \multicolumn{1}{c}{} & \multicolumn{1}{c}{} & \multicolumn{1}{c}{[deg]} & \multicolumn{1}{c}{[deg]} & \multicolumn{1}{c}{[days]} & \multicolumn{1}{c}{[mag]} & \multicolumn{1}{c}{dex}\\
        \hline
            NGC1904 & V4 & 0	&  H99 &	81.074 & -24.538 &	0.634 &	0.78 $\pm$ 0.06 & -1.60 \\
            NGC1904	& V13 & 0	&  H99 &	81.044	& -24.520 &	0.689 &	0.72 $\pm$ 0.07	& -1.60 \\
            NGC5634	& V2 & 0	&  H99 &	217.399	& -5.954 &	0.605 &	1.30 $\pm$ 0.20	& -1.88 \\
            NGC5634	& V3 & 0	&  H99 &	217.393	& -5.965 &	0.601 &	0.83 $\pm$ 0.11	& -1.88 \\
            NGC6235	& V1 & 1	&  G-E &	253.351	& -22.167 &	0.616 &	0.91 $\pm$ 0.03	& -1.28 \\
        \hline
    \end{tabular}
\tablefoot{Name and Var are the names of the GGC hosting the RRLs and the RRL identification according to \citet{clement2001variable}; classification of the GGCs according to \citet{belokurov2024situ}: \textit{In-Situ}=1 \textit{Accreted}=0; classification of GGCs according to \citet{callingham2022chemo}, the meaning of the acronyms is explained in Table~\ref{tab:prog}; RA and DEC are the equatorial coordinates (J2000) according to \citet{clement2001variable}; Period and Amp(G) are from $Gaia$ DR3 \citep[][]{Clementini2023}; [Fe/H] is from \citet{Harris1996}. Only a portion of the table is listed here to guide the reader about its content. A machine-readable version of the full table will be published at the CDS (Centre de Données astronomiques de Strasbourg, https://cds.u-strasbg.fr/} 
\end{table*}

\begin{figure*}[htbp]
    \centering
    \includegraphics[width=1\textwidth, height=1\textheight, keepaspectratio]{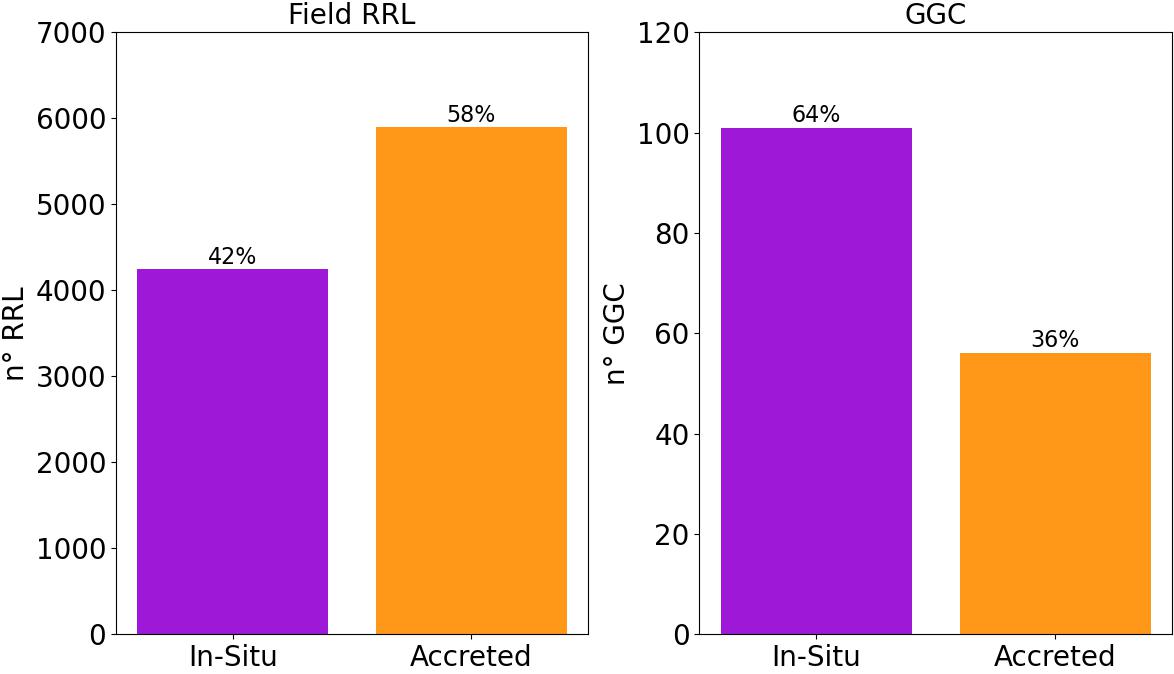}
    \caption{\textit{Left:} Bar chart of the field RRL sample where dark violet and dark orange indicate the \textit{In-situ} and the \textit{Accreted} population, respectively. \textit{Right}: as before but for the GGCs.}
        \label{Fig:in_situ_accreted_bar_chart}
\end{figure*}

\begin{figure*}[htbp]
    \centering
    \includegraphics[width=1\textwidth, height=1\textheight, keepaspectratio]{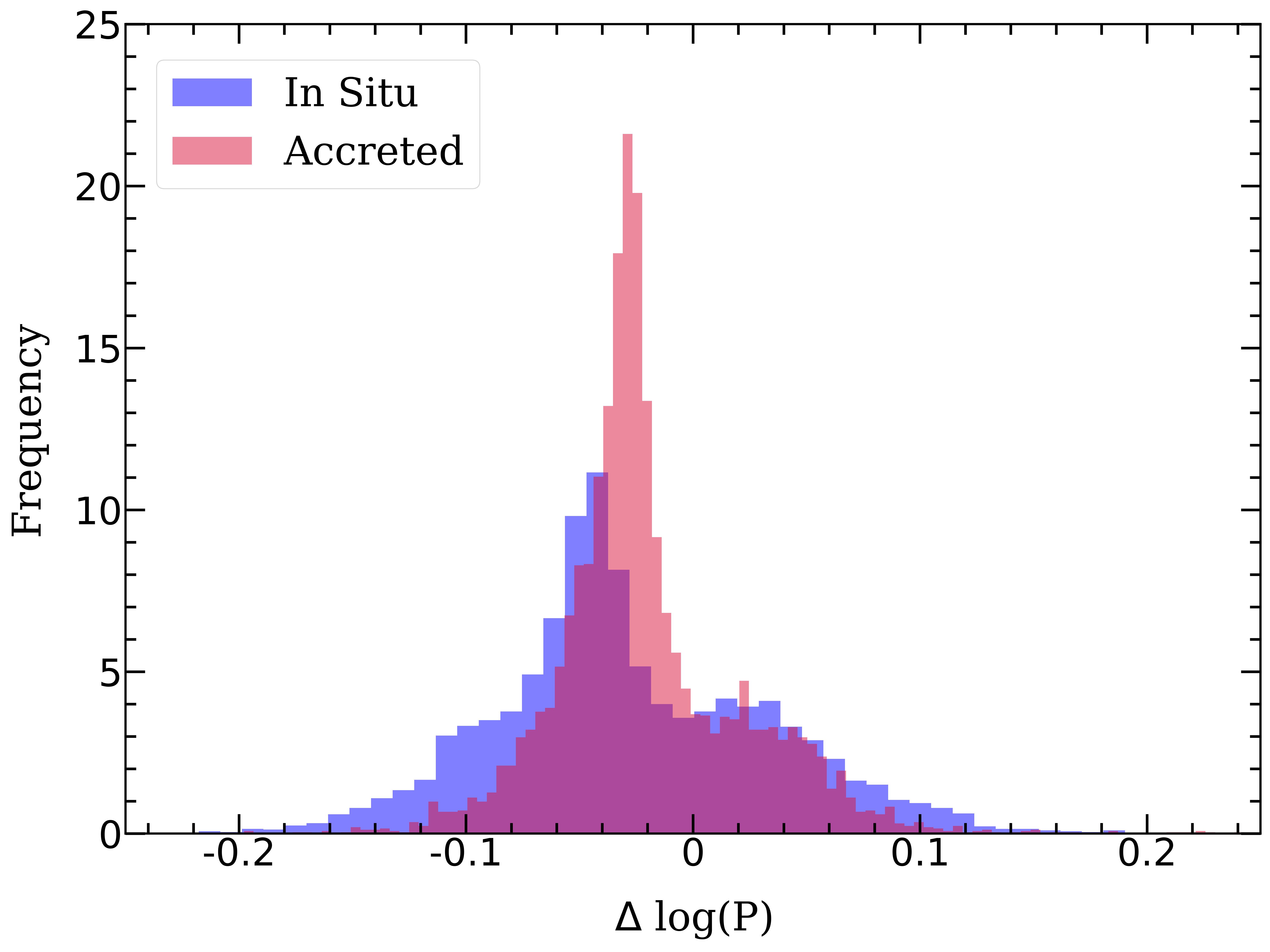}
    \caption{"Rectification" of the Bailey diagram. The histogram shows the distribution of $\Delta\,log\,(P)=log\,(P) - log\,(P_{func})$, where "func" is the line (polynomial function) separating the Oosterhoff types shown in Fig.~\ref{Fig:Bailey_GGC_field_RRL}. The \textit{In-Situ} and \textit{Accreted} populations are shown in blue and red, respectively. 
        \label{Fig:delta_log_p}}
\end{figure*}

\begin{figure*}[htbp]
    \centering
    \includegraphics[width=1\textwidth, height=1\textheight, keepaspectratio]{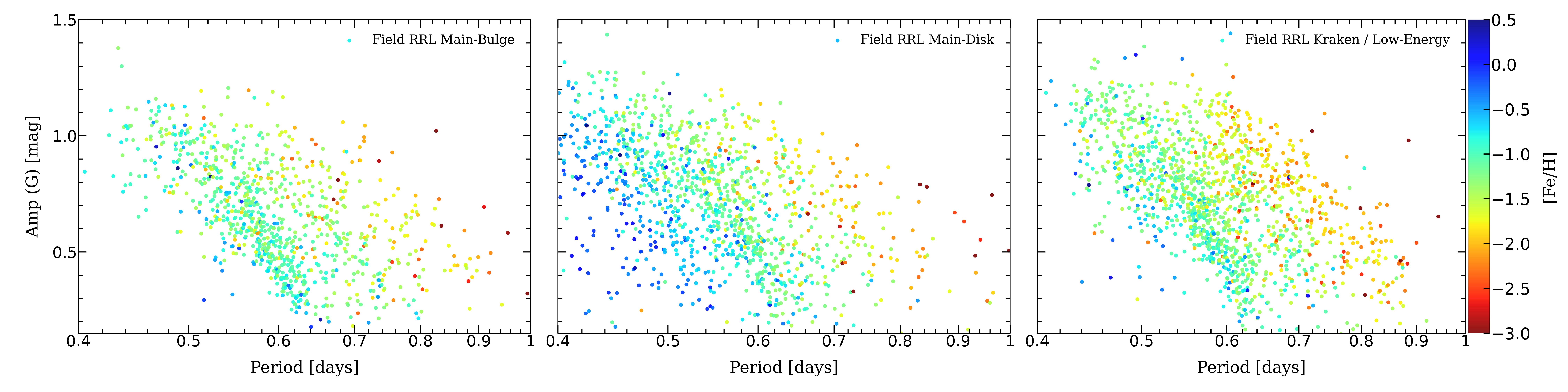}
    \caption{Bailey diagram of field RRLs associated with the \textit{In-Situ} population, separated according to the \textit{Progenitors}' classification by \citet{callingham2022chemo} (see labels). Sources are colour-coded according to their metallicity.}
        \label{Fig:Bailey_field_RRL_prog_in_situ}
\end{figure*}

\begin{figure*}[htbp]
    \centering
    \includegraphics[width=1\textwidth, height=1\textheight, keepaspectratio]{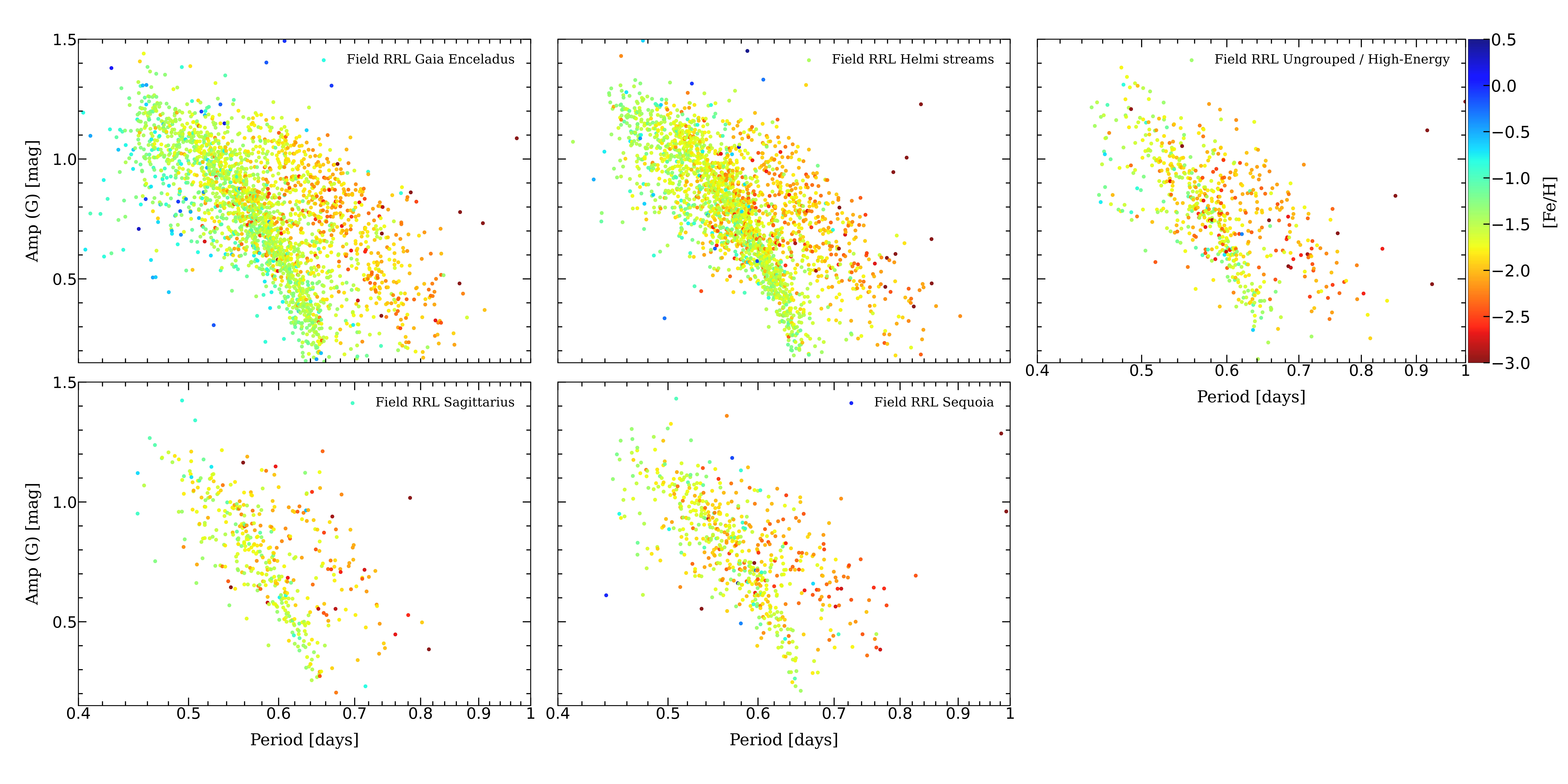}
    \caption{Bailey diagram of field RRLs associated with the \textit{Accreted} population, separated according to the \textit{Progenitors}' classification by \citet{callingham2022chemo} (see labels). Sources are colour-coded according to their metallicity.}
        \label{Fig:Bailey_field_RRL_prog_acc}
\end{figure*}

\end{appendix}

\end{document}